\documentclass[a4paper]{jpconf}
\usepackage{graphicx}
\usepackage{bm}
\usepackage{braket}
\usepackage{color}
\usepackage{cite}
\begin{document}
\title{Competition between Quadrupole and Magnetic Kondo Effects in Non-Kramers Doublet Systems}

\author{Hiroaki Kusunose$^{1}$, Takahiro Onimaru$^{2}$}

\address{$^{1}$ Department of Physics, Ehime University, Matsuyama, 790-8577, Japan}


\address{$^{2}$ Graduate School of AdSM, Hiroshima University, Higashi-Hiroshima, 739-8530, Japan}

\ead{hk@ehime-u.ac.jp}

\begin{abstract}
We discuss possible competition between magnetic and quadrupole Kondo effects in non-Kramers doublet systems under cubic symmetry. The quadrupole Kondo effect leads to non-Fermi-liquid (NFL) ground state, while the magnetic one favors ordinary Fermi liquid (FL). In terms of the $j$-$j$ coupling scheme, we emphasize that the orbital fluctuation must develop in the vicinity of the NFL-FL boundary.
We demonstrate a change of behavior in the f-electron entropy by the Wilson's numerical renormalization-group (NRG) method on the basis of the extended two-channel Kondo exchange model.
We present implications to extensively investigated PrT$_{2}$X$_{20}$ (T=Ti, V, Ir; X=Al, Zn) systems that exhibit both quadrupole ordering and peculiar superconductivity. We also discuss the magnetic-field effect which lifts weakly the non-Kramers degeneracy. Our model also represents the FL state accompanied by a free magnetic spin as a consequence of stronger competition between the magnetic and the quadrupole Kondo effects.
\end{abstract}

\section{Introduction}

In contrast to the standard heavy-fermion physics so far developed extensively in Ce-based compounds, praseodymium (and uranium) materials have more exotic aspects inherent in a variety of their atomic states.
Noz\`eires and Blandin first argued this issue in view of Kondo effect, and they categorized possible ground states (fixed points) in terms of the magnitude of spin $S$ and the number of scattering channels of conduction electrons $n_{c}$ as (i) the perfect screening ($n_{c}=2S$), (ii) the underscreening ($n_{c}<2S$), and (iii) the overscreening ($n_{c}>2S$)~\cite{Nozieres80}.

Among them, the overscreening case results in the NFL phenomena, and extensive investigations on this subject have been performed.
For instance, the temperature dependences of the specific heat, magnetic susceptibility, and resistivity for $n_{c}=2$ and $S=1/2$ are given by
\begin{equation}
C(T)/T,\,\,\chi(T)\propto
-\ln(T/T_{\rm 2K}),
\quad
\rho(T)-\rho_{0}\propto (T/T_{\rm 2K})^{1/2},
\end{equation}
and the residual entropy is given by $S(0)=(1/2)\ln 2$,
where $T_{\rm 2K}$ is the Kondo temperature for the two-channel Kondo exchange model~\cite{Cox98,Emery92,Affleck95,Tsvelick85,Sacramento89,Sakai93,Cox93}, and $\rho_{0}$ is the residual resistivity.
In the last decade, the lattice version of the two-channel Kondo problem has also been studied theoretically~\cite{Jarrell96}, in particular, it is found by means of the so-called dynamical mean-field theory that it exhibits a variety of long-range orderings~\cite{Hoshino13,Hoshino11}, and analytic expressions of $T$ dependence are obtained such as
\begin{equation}
C(T)/T,\chi\propto-(T/T_{\rm F2}^{*})^{1/2},
\quad
\rho(T)-\rho_{0}\propto(1+T/T_{\rm F2}^{*})^{-1},
\end{equation}
in the $1/N$ expansion with vertex corrections~\cite{Tsuruta00,Tsuruta14}.
$T_{\rm F2}^{*}$ is the characteristic temperature of the two-channel Kondo lattice model, which differs from the single-site counterpart $T_{\rm 2K}$ in general.

A practical situation to realize the overscreening was proposed by Cox~\cite{Cox98,Cox87,Cox93a}.
In his model for UBe$_{13}$, the quadrupole moment in the non-Kramers doublet plays a role of pseudospin ($S=1/2$), which is to be screened by the quadrupole moment of conduction electrons with real spins, $\uparrow$ and $\downarrow$, as two equivalent scattering channels ($n_{c}=2$).
Hence, it is called the quadrupole Kondo effect.
In the Uranium ion, $5f$ electron is moderately localized and the crystalline-electric-field (CEF) splitting is often too broad to identify it.
This aspect has hampered clear understanding of such exotic Kondo effect.
On the contrary, $4f$ electron in the Pr ion is well localized exhibiting clear CEF splitting.
Moreover, a cage-like structure such as the filled Skutterudites provides many number of ligands for f electrons, which effectively increases hybridization strength to conduction electrons~\cite{Sato09}.
Such a situation offers ideal playground to explore the exotic Kondo problem and resultant anomalous electronic states.

Recently, a series of PrT$_{2}$X$_{20}$ compounds, having a cage-like structure as well, has been found as non-Kramers doublet system, which meets criteria for the quadrupole Kondo effect.
In fact, PrV$_{2}$Al$_{20}$, PrIr$_{2}$Zn$_{20}$ and PrRh$_{2}$Zn$_{20}$ exhibit NFL behaviors in the resistivity, specific heat and so on above the quadrupole ordering~\cite{Sakai11,Onimaru10,Onimaru12}.
They even coexist with superconductivity~\cite{Tsujimoto14,Onimaru11,Onimaru12}.
On the contrary, PrTi$_{2}$Al$_{20}$ shows ordinary FL behavior at ambient pressure~\cite{Sakai11}.
From the theoretical point of view, a simple construction of the extended two-channel Kondo exchange model with non-Kramers doublet CEF ground state under cubic symmetry does not possess any FL fixed points~\cite{Koga99}, although the model without CEF splitting must apparently have FL fixed point.

In this paper, we construct yet another extended two-channel Kondo exchange model, which represents the competition between NFL and FL.
These two fixed points differ in the occupation of two f electrons onto $f^{1}$ orbitals, and hence the orbital fluctuation must be important in the vicinity of the boundary between these two fixed points.
The organization of the paper is as follows.
In the next section, we introduce the extended two-channel Kondo exchange model and discuss the limiting cases.
In \S3, we exhibit results obtained by Wilson's NRG method.
We discuss the distinct electronic states and their parameter dependence in the characteristic temperatures.
The magnetic-field effect is also discussed.
In the last section, we discuss relevance of our results to recent experiments in PrT$_{2}$X$_{20}$ systems, and summarize the paper.

\section{Extended Two-Channel Kondo Model}

\subsection{The $f^{2}$ CEF states in the $j$-$j$ coupling scheme}

With an odd number of f-electron valency, localized CEF states are subject to the Kramers theorem, which states that at least double degeneracy must remain due to time-reversal symmetry.
On the contrary, systems with an even number of f electrons are free from this theorem, and a degeneracy should come from different symmetry reason other than the time reversal, or may be accidental.

For example, the total angular momentum $J=4$ multiplet of $f^{2}$ configuration is lifted into $\Gamma_{1}$ singlet, $\Gamma_{3}$ doublet, and $\Gamma_{4}$, $\Gamma_{5}$ triplets under cubic symmetry.
Among them, $\Gamma_{3}$ doublet carries no magnetic moments, and it has $(3z^{2}-r^{2})$-type and $(x^{2}-y^{2})$-type electric quadrupoles, in addition to $(J_{x}J_{y}J_{z})$-type magnetic octupole.
The wavefunctions of these CEF states are complicated and rather depend on whether the $LS$ coupling (Russell-Saunders) or the $j$-$j$ coupling scheme is used.
Nevertheless, essential feature of the states can be understood in a simplified version of the $j$-$j$ coupling picture as follows.
More precise expressions of the wavefunctions in the $j$-$j$ coupling scheme can be found in the literature~\cite{Kusunose05}.

\begin{figure}
\begin{center}
\includegraphics[width=15.8cm]{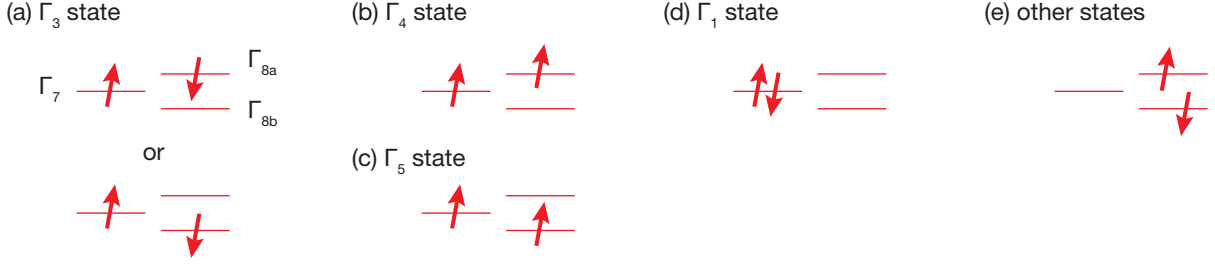}
\caption{\label{f2cef}The $f^{2}$ CEF states in the $j$-$j$ coupling scheme. (a) $\Gamma_{3}$ doublet, (b,c) $\Gamma_{4}$, $\Gamma_{5}$ triplets, (d) $\Gamma_{1}$ singlet, and (e) other states.}
\end{center}
\end{figure}

In the $j$-$j$ coupling scheme, the $f^{2}$ states are described by filling up two electrons into $f^{1}$ CEF states, which are $\Gamma_{7}$ doublet, and $\Gamma_{8}$ quartet under cubic symmetry.
The three Kramers pairs (labelled by the orbital indices, $m=7$, $a$, $b$) constitute $\Gamma_{7}$ and $\Gamma_{8}$ states, and the latter is degenerate due to cubic symmetry.
We neglect the energy splitting between $\Gamma_{7}$ and $\Gamma_{8}$ for simplicity.
Then, the non-Kramers $\Gamma_{3}$ states can be regarded as two singlet states of $\Gamma_{7}$-$\Gamma_{8a}$ and $\Gamma_{7}$-$\Gamma_{8b}$, while the two magnetic triplet states, $\Gamma_{4}$, $\Gamma_{5}$ correspond to two triplets among $\Gamma_{7}$-$\Gamma_{8}$.
We should note that in these states, one f electron always occupies $\Gamma_{7}$ and another stays in either $\Gamma_{8a}$ or $\Gamma_{8b}$ state.
The remaining $\Gamma_{1}$ singlet is identified as the double-occupied state within $\Gamma_{7}$ state.
The role of the orbital degeneracy is apparent in the above construction of the $f^{2}$ CEF states, as shown schematically in Fig.~\ref{f2cef}.

The CEF splitting measured from the $\Gamma_{3}$ CEF ground state can be expressed by
\begin{equation}
H_{f}=\Delta(P_{fa}+P_{fb})+\Delta_{1}\,n_{fa}\,n_{fb},
\quad
P_{fm}\equiv\bm{S}_{fm}\cdot\bm{S}_{f7}+\frac{3}{4}n_{fm}\,n_{f7},
\end{equation}
where
\begin{equation}
\bm{S}_{fm}=\sum_{\alpha\beta}f_{m\alpha}^{\dagger}\left(\frac{\bm{\sigma}_{\alpha\beta}}{2}\right)f_{m\beta}^{},
\quad
n_{fm}=\sum_{\alpha}f_{m\alpha}^{\dagger}f_{m\alpha}^{},
\end{equation}
are the spin and number operators for each $f^{1}$ Kramers pair, and $P_{fm}$ is the projection operator onto the $\Gamma_{8m}$-$\Gamma_{7}$ triplet state.
We have neglected the doubly occupied states in each $f^{1}$ orbital such as Fig.~\ref{f2cef}(d) due to large intra Coulomb repulsion, and the splitting between $\Gamma_{4}$ and $\Gamma_{5}$ excited states, for simplicity.
Moreover, we have introduced other energy scale $\Delta_{1}$ for the excited states such as Fig.~\ref{f2cef}(e).

\subsection{Magnetic and orbital exchange couplings}

When the hybridization $(V_{8}\sum_{km\alpha}f_{m\alpha}^{\dagger}c_{km\alpha}^{}+{\rm h.c.})$ between the f and corresponding partial wave of conduction electrons is switched on, magnetic and orbital exchange processes appear in the second order of $V_{8}$.
In terms of the Pauli matrices, $\bm{\tau}$, acting on the (a,b) orbital space, the orbital operator for f electrons is given by
\begin{equation}
\bm{T}_{f}=\sum_{\alpha}\sum_{mn}^{a,b}f_{m\sigma}^{\dagger}\left(\frac{\bm{\tau}_{mn}}{2}\right)f_{n\alpha}^{}.
\end{equation}
Then, the two exchange processes are expressed as
\begin{equation}
H_{\rm ex}=J\bigl[\bm{s}_{a}(0)\cdot\bm{S}_{fa}+\bm{s}_{b}(0)\cdot\bm{S}_{fb}\bigr]+K\bigl[\bm{t}_{\uparrow}(0)+\bm{t}_{\downarrow}(0)\bigr]\cdot\bm{T}_{f},
\end{equation}
where $\bm{s}_{m}(0)$ and $\bm{t}_{\alpha}(0)$ are the spin and orbital operators of the conduction electrons at the impurity site.
Here, $\bm{T}_{f}$ couples equally with two orbital degrees of freedom, $\bm{t}_{\alpha}(0)$ ($\alpha=\uparrow$ and $\downarrow$ channels), which is ensured by the time-reversal symmetry.
The second term is the origin of the overscreening Kondo effect of quadrupole degrees of freedom as will be shown shortly.

Since the origins of two exchange processes are the same hybridization $V_{8}$, $J$ and $K$ are expected to be the same order of magnitude, i.e., $J,K\sim |V_{8}|^{2}/\Delta E$, where $\Delta E$ is the energy difference between $f^{2}$ and $f^{1,3}$ states.
In contrast to $f^{1}$ systems such as Ce-based compounds where magnetic Kondo effect usually dominates, the hybridization leads to both magnetic and quadrupole Kondo effects, especially in the non-Kramers doublet systems with suitable magnetic excited states under cubic symmetry.
Therefore, an interesting competition between $J$ and $K$ terms could be expected in those systems.
We will ignore the hybridization among $\Gamma_{7}$ as a first step.

Eventually, the Hamiltonian of the extended two-channel Kondo model to be investigated in this paper reads
\begin{equation}
H=\sum_{km\alpha}\epsilon_{k}c_{km\alpha}^{\dagger}c_{km\alpha}+H_{\rm ex}+H_{f}.
\label{model}
\end{equation}

\subsection{Two limiting cases}

Let us consider two opposite limits of the model.
First, we consider the case where the excited CEF states can be neglected, or $K\gg J$.
Within the $\Gamma_{3}$ doublet denoted by $\ket{u}$ and $\ket{v}$, $\bm{S}_{fm}\to0$ and
\begin{equation}
\bm{T}_{f}\to \bm{\tau}_{f}\equiv\sum_{\mu\nu}^{u,v}f_{\mu}^{\dagger}\left(\frac{\bm{\tau}_{\mu\nu}}{2}\right)f_{\nu}^{}.
\end{equation}
Then, our model is reduced to the ordinary two-channel Kondo model as
\begin{equation}
H=\sum_{\alpha}^{\uparrow,\downarrow}\left[\sum_{k}\sum_{m}^{a,b}\epsilon_{k}c_{km\alpha}^{\dagger}c_{km\alpha}^{}+K\bm{t}_{\alpha}(0)\cdot\bm{\tau}_{f}\right],
\end{equation}
where $\alpha$ plays a role of two exchange channels, while $m$ or $\mu$ acts as the pseudo spins.
Therefore, we expect the NFL ground state in this limit.

The other limit is $J\gg K$, $\Delta_{1}$.
In this case, both f electrons tend to occupy $\Gamma_{8}$ states like Fig.~\ref{f2cef}(e) in order to gain the magnetic exchange energy, $J$.
In this case, $\bm{S}_{f7}$ and $\bm{T}_{f}$ become inactive, and the model is reduced to
\begin{equation}
H=\sum_{m}^{a,b}\left[\sum_{k}\sum_{\alpha}^{\uparrow,\downarrow}\epsilon_{k}c_{km\alpha}^{\dagger}c_{km\alpha}^{}+J\bm{s}_{m}(0)\cdot\bm{S}_{fm}\right].
\end{equation}
This is nothing but two independent single-channel Kondo models, showing ordinary FL behaviors.

\subsection{Magnetic-field effect}

Since the non-Kramers doublet has no magnetic dipoles, the magnetic-field effect appears in the order of $h^{2}/\Delta$ via the excited magnetic states.
In order to describe such a situation for $\bm{h}\parallel z$, we introduce the following operator as a ``magnetic'' moment,
\begin{equation}
M_{z}=\sum_{m}^{7,a,b}S_{fm}^{z}+\frac{1}{2}(g_{1}K_{f1}+g_{2}K_{f2}),
\quad
K_{fm}\equiv n_{fm\uparrow}n_{f7\downarrow}-n_{fm\downarrow}n_{f7\uparrow},
\end{equation}
where $|g_{1}|\ne|g_{2}|$.
In what follows, we take $g_{1}=1$, $g_{2}=0$ for simplicity.
Including the Zeeman coupling $H_{Z}=-hM_{z}$, the $f^{2}$ CEF states are summarized in Table~\ref{tbl1}.

\begin{table}
\caption{\label{tbl1}The $f^{2}$ CEF states under magnetic field along $z$ axis.}
\begin{center}
\begin{tabular}{cccclc}
\br
$S_{f}^{z}$ & deg. & energy (E) & E ($h=0$) & role & eigenstate for $h=0$ \\
\mr
$0$ & 1 & $\frac{\Delta}{2}-\frac{\Delta}{2}\sqrt{1+(g_{1} h/\Delta)^{2}}$ & $0$ & $\Gamma_{3}$ & $\ket{u}\equiv\frac{1}{\sqrt{2}}(\ket{a,\uparrow}\ket{7,\downarrow}-\ket{a,\downarrow}\ket{7,\uparrow})$ \\
& 1 & $\frac{\Delta}{2}-\frac{\Delta}{2}\sqrt{1+(g_{2} h/\Delta)^{2}}$ & & & $\ket{v}\equiv\frac{1}{\sqrt{2}}(\ket{b,\uparrow}\ket{7,\downarrow}-\ket{b,\downarrow}\ket{7,\uparrow})$ \\
\mr
$-1$ & 2 & $\Delta+h$ & $\Delta$ & $\Gamma_{4,5}$ & $\ket{a,\downarrow}\ket{7,\downarrow}$, $\ket{b,\downarrow}\ket{7,\downarrow}$ \\
$0$ & 1 & $\frac{\Delta}{2}+\frac{\Delta}{2}\sqrt{1+(g_{1} h/\Delta)^{2}}$ & & & $\frac{1}{\sqrt{2}}(\ket{a,\uparrow}\ket{7,\downarrow}+\ket{a,\downarrow}\ket{7,\uparrow})$ \\
$0$ & 1 & $\frac{\Delta}{2}+\frac{\Delta}{2}\sqrt{1+(g_{2} h/\Delta)^{2}}$ & & & $\frac{1}{\sqrt{2}}(\ket{b,\uparrow}\ket{7,\downarrow}+\ket{b,\downarrow}\ket{7,\uparrow})$ \\
$+1$ & 2 & $\Delta-h$ & & & $\ket{a,\uparrow}\ket{7,\uparrow}$, $\ket{b,\uparrow}\ket{7,\uparrow}$ \\
\mr
$-1$ & 1 & $\Delta_{1}+h$ & $\Delta_{1}$ & others & $\ket{a,\downarrow}\ket{b,\downarrow}$ \\
$0$ & 2 & $\Delta_{1}$ & & & $\ket{a,\uparrow}\ket{b,\downarrow}$, $\ket{a,\downarrow}\ket{b,\uparrow}$ \\
$+1$ & 1 & $\Delta_{1}-h$ & & & $\ket{a,\uparrow}\ket{b,\uparrow}$ \\
\br
\end{tabular}
\end{center}
\end{table}

\section{Results by Wilson's NRG Calculation}

We have applied the Wilson's NRG method~\cite{Wilson75,Bulla08} to solve our model, (\ref{model}).
In this numerical method, the conduction band is discretized logarithmically to focus on the low-energy excitations.
To do so, the kinetic-energy term is transformed to the hopping-type Hamiltonian on the semi-infinite one-dimensional chain, which can be solved by iterative diagonalization.
In this paper, we have used the discretization parameter $\Lambda=3$, and retained at least $N_{\rm r}=700$ low-energy states for each iteration step.
The CEF parameters are chosen as $\Delta=1.0\times 10^{-3}$ and $\Delta_{1}=5.0\times 10^{-3}$ in unit of the half-bandwidth of the conduction electron, $D$.
We can keep track of the low-energy states by the conserved quantities, $\Phi=$($Q$, $n_{f7}$, $S_{z}$), where $Q$ is the total number of electrons measured from half filling, $n_{f7}$ is the number of f electrons in the $\Gamma_{7}$ orbital, and $S_{z}$ is the $z$-component of the total spin.
The conserved quantities only for even-number of iteration step, $N$, will be given, if necessary.

\subsection{Fixed points}

\begin{figure}
\begin{center}
\includegraphics[width=15.8cm]{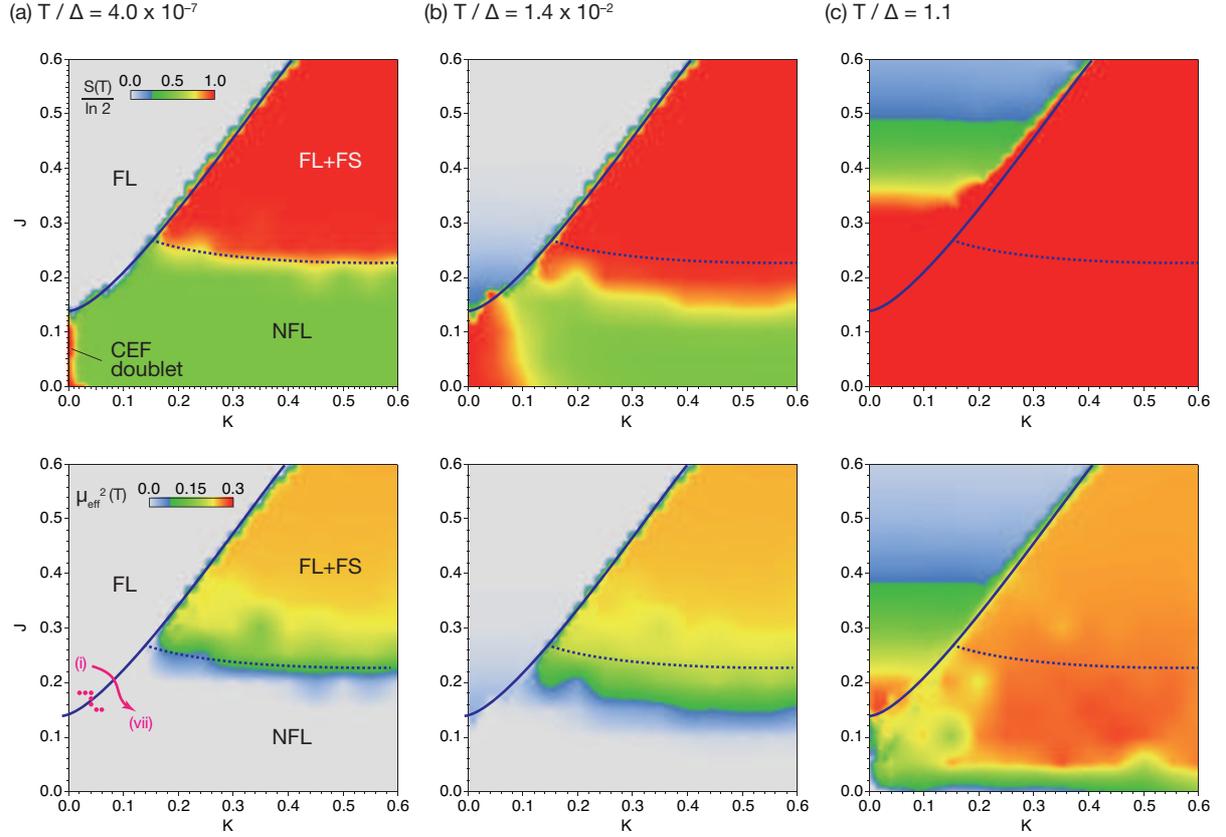}
\caption{\label{s-m2}The $T$ dependences of the f-electron entropy (upper panels) and the squared effective moment (lower panels) for (a) $T/\Delta=4.0\times 10^{-7}$, (b) $T/\Delta=1.4\times 10^{-2}$, and (c) $T/\Delta=1.1$. At the lowest $T$, the local FL state, NFL state, and FL plus a marginally free spin $S=1/2$ (FL+FS) state exist in $J$-$K$ space. The boundary between the FL and NFL/FL+FS (solid line) is the first order, while that between NFL and FL+FS (dotted line) is a crossover. For small $J$ and $K$, the unscreened CEF doublet state remains at the lowest $T$. The arrow with 7 points in the lower left panel indicates the parameter sets used for the later discussions.}
\end{center}
\end{figure}

\begin{figure}
\begin{center}
\includegraphics[width=14cm]{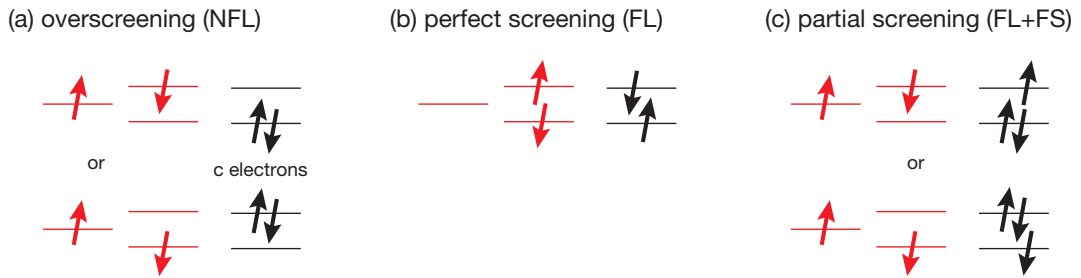}
\caption{\label{fp}Schematic picture of three distinct low-energy states, (a) the overscreening (NFL), (b) the perfect screening (FL), and (c) the partial screening (FL+FS). The FL and NFL states have no magnetic moments, while the marginally free spin remains in the FL+FS state.}
\end{center}
\end{figure}

Figure~\ref{s-m2} shows $T$ dependences of the f-electron entropy $S(T)$, and its squared effective moment, $\mu_{\rm eff}^{2}(T)$.
There are three distinct states at the lowest $T$ as shown in Fig.\ref{s-m2}(a), two of which are nothing but the states we expected in the limiting cases.
Namely, the local FL state $\Phi=(0,0,0)$ appears for $J>K$, while the local NFL state $\Phi=(0,1,0)$ for $J<K$.
In addition to them, the FL plus a marginally free spin (FL+FS) state $\Phi=(1,1,\pm1/2)$ with $\mu_{\rm eff}^{2}=1/4$ appears for $J\approx K$.
The FL+FS state emerges as a consequence of the competition between the magnetic and quadrupole Kondo effects, in which one additional electron (for even $N$) or hole ($Q=-1$ for odd $N$) is bound around the impurity site.
The schematic picture for these states is shown in Fig.~\ref{fp}.
For small $J$ and $K$, the unscreened CEF doublet state remains at the lowest $T$, since the characteristic energies of the quadrupole and magnetic Kondo effects, $T_{\rm 2K}$ and $T_{\rm K}$ are much below the accessible $T$.

The boundary between FL and NFL/FL+FS is the first order as $n_{f7}$ changes at the boundary.
The critical value of $J$ at $K=0$ is determined by the condition that $T_{\rm K}\approx \Delta_{1}$.
On the other hand, the boundary between NFL and FL+FS is crossover.
As we approach from the NFL state to the FL+FS state, $T_{\rm 2K}$ decreases considerably, and the system eventually stays in the FL+FS state over the calculated $T$ range.

\subsection{FL-NFL transition}

\begin{figure}
\begin{center}
\includegraphics[width=15.8cm]{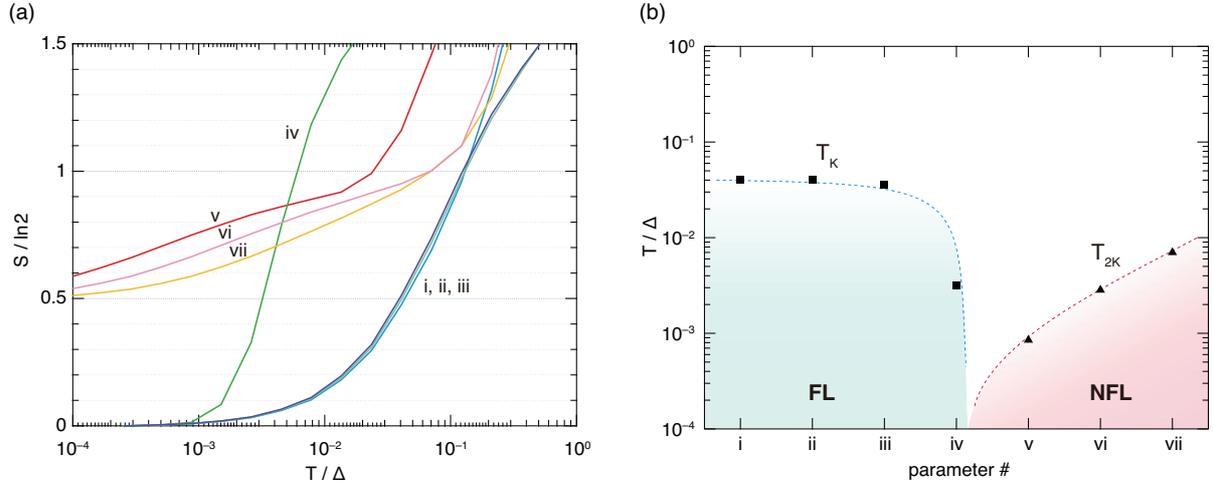}
\caption{\label{flnfl}(a) The change of the entropy across the FL-NFL boundary. The parameter set is given by (i) $J=0.18$, $K=0.02$, (ii) $J=0.18$, $K=0.03$, (iii) $J=0.18$, $K=0.04$, (iv) $J=0.17$, $K=0.04$, (v) $J=0.16$, $K=0.04$, (vi) $J=0.15$, $K=0.05$, and (vii) $J=0.15$, $K=0.06$ as indicated by the arrow with 7 points in Fig.~\ref{s-m2}(a), and (b) the Kondo temperatures, $T_{\rm K}$ and $T_{\rm 2K}$ for the FL and NFL states. The dotted lines are guide to eye.}
\end{center}
\end{figure}

Let us discuss the transition between the FL and NFL states.
The $T$ dependence of the entropy is shown in Fig.~\ref{flnfl}(a).
The parameter sets used for the calculation are indicated by the arrow with 7 points in Fig.~\ref{s-m2}(a).
The magnetic Kondo temperature $T_{\rm K}$ has weak parameter dependence except the vicinity of the boundary.
On the contrary, the quadrupole Kondo temperature $T_{\rm 2K}$ is rather sensitive to the parameters.
The quantitative parameter dependences of $T_{\rm K}$ and $T_{\rm 2K}$ are shown in Fig.~\ref{flnfl}(b), where $T_{\rm K}$ ($T_{\rm 2K}$) is defined by $S(T_{\rm K})=0.5\ln 2$ ($S(T_{\rm 2K})=0.75\ln 2$).
In the vicinity of the boundary, the both characteristic energies become extremely low, reflecting a level-crossing nature of two states.
As was shown in Fig.~\ref{fp}(a) and (b), the f-electron occupancy differs in two states, and hence the strong orbital fluctuation between $\Gamma_{7}$ and $\Gamma_{8}$ must be expected near the FL-NFL boundary.
This is the important consequence of the cubic symmetry and the non-Kramers CEF ground state.

\subsection{Magnetic-field effect}

\begin{figure}
\begin{center}
\includegraphics[width=15.8cm]{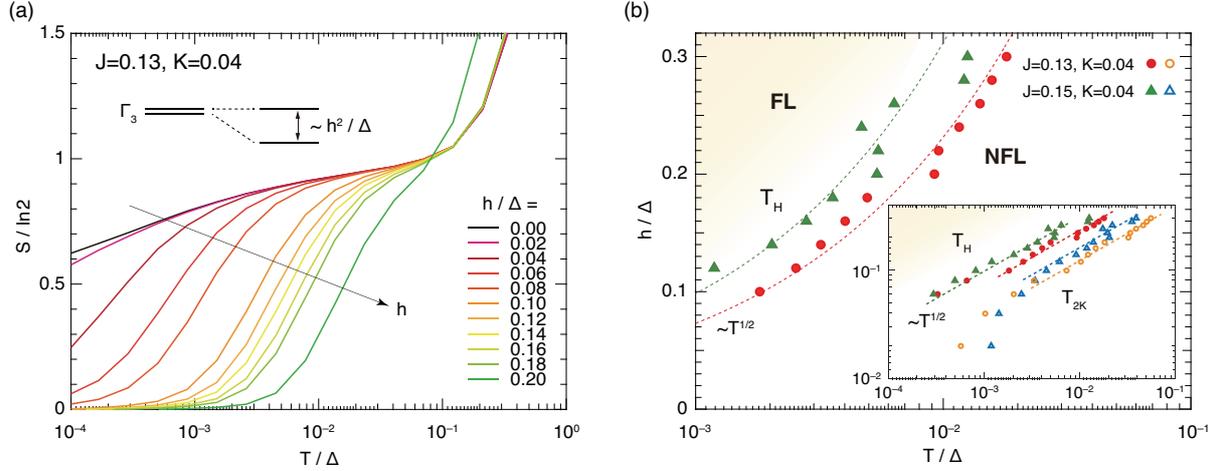}
\caption{\label{sh}The magnetic-field effect on the NFL state in (a) the entropy, and (b) the crossover temperature to the FL state $T_{\rm H}$, which shows non-monotonous field dependences in stronger fields. Such non-monotonous tendency becomes more prominent as the FL-NFL boundary is approached. The inset shows that $T_{\rm H}$ and the quadrupole Kondo temperature $T_{\rm 2K}$ roughly follow $\propto h^{2}$, although $T_{\rm H}\propto h^{4}$ is predicted for the two-channel Kondo model in the weak-field limit.}
\end{center}
\end{figure}

We discuss the magnetic-field effect along the $z$ axis.
The non-Kramers doublet is lifted by the field through the excited magnetic states.
Thus, the magnitude of the splitting is proportional to $h^{2}/\Delta$ in the weak-field limit as shown in Table~\ref{tbl1}.
This splitting brings about inequivalency of two channels in the model, yielding a crossover from the NFL to the FL state of an effective singlet-channel Kondo problem.
Note that this FL state is completely different from the FL state by the magnetic Kondo effect as shown in Fig.~\ref{fp}(b).
According to the scaling dimensional analysis for the two-channel Kondo model, the crossover scale $T_{\rm H}$ is proportional to square of the energy splitting, i.e., $T_{\rm H}\propto(h^{2}/\Delta)^{2}/T_{\rm 2K}$~\cite{Yotsuhashi02}.
However, as the field increases, there must be a discrepancy from the scaling analysis owing to the presence of the excited CEF states and the transition between the NFL and FL states.

Figure~\ref{sh} shows the $h$ dependence of the entropy and the relevant energy scales, $T_{\rm H}$ and $T_{\rm 2K}$, for the NFL state ($J=0.13$, $K=0.04$), where the $T_{\rm H}$ is defined by $S(T_{\rm H})=0.25\ln2$.
Although the $T_{\rm H}$ indeed follows $h^{2}$ in very weak-field range (not shown), it is roughly fitted by $T_{\rm H}\propto h^{2}$ above $h>0.1\Delta$ as shown by the dotted lines in Fig.~\ref{sh}(b) and the inset.
The quadrupole Kondo temperature $T_{\rm 2K}$ also follows $\propto h^{2}$ in the same field range.
Furthermore, $T_{\rm H}$ shows non-monotonous field dependences in stronger fields.
Such non-monotonous tendency becomes more prominent as the FL-NFL boundary is approached.

The value of the magnetization for the NFL state is strongly suppressed owing to the non-Kramers CEF character.
If a system is close enough to the NFL-FL boundary, the magnetic field causes a transition to the FL state of the magnetic Kondo effect.
It is a transition from the field-induced FL to the magnetic Kondo driven FL accompanied by the change of the orbital occupation from $n_{f7}=1$ to $n_{f7}=0$.
In this case, the field energy easily exceeds $T_{\rm K}$ (see Fig.~\ref{flnfl}(b)), yielding larger magnetization value.
Thus, the field-induced transition to the FL state exhibits a metamagnetic behavior.

\section{Discussions and Summary}
Let us now discuss relevance of our results to recent experiments although our model takes account of only single-site Kondo effects.
PrT$_{2}$X$_{20}$ (e.g., T=V, Ti, Ir, Rh; X=Al, Zn) systems commonly have $\Gamma_{3}$ non-Kramers CEF ground state, and the first excited magnetic states $\Gamma_{4}$ or $\Gamma_{5}$.
For example, in PrIr$_{2}$Zn$_{20}$ the excited $\Gamma_{4}$ is located at 27 K~\cite{Iwasa13}.
Both PrTi$_{2}$Al$_{20}$ and PrV$_{2}$Al$_{20}$ exhibit the logarithmic $T$ dependence in the 4f-electron contribution of the resistivity above 70 K, indicating that magnetic Kondo effect occurs in higher temperature range.
However, the latter shows a NFL behavior with $\sim T^{1/2}$ below $30$ K, while the former shows FL $T^{2}$ dependence below 20 K~\cite{Sakai11} as long as they are in the paramagnetic phase.
PrTi$_{2}$Al$_{20}$ (PrV$_{2}$Al$_{20}$) undergoes a transition to the ferro (antiferro) quadrupole ordering at $T_{\rm Q}=2$ K (0.6 K), and then to the superconducting state at $T_{c}=0.2$ K (0.05 K for the high-quality sample)~\cite{Sakai12,Tsujimoto14}.

Applying pressure, the resistivity in PrTi$_{2}$Al$_{20}$ shows a crossover from the FL behavior to the NFL behavior around $P\sim 3$ GPa, and further increase of pressure, $T_{\rm Q}$ decreases slightly while $T_{\rm F2}^{*}$ considerably increases above $10$ GPa.
$T_{c}$ shows a peak structure at the lowest $T_{\rm F2}^{*}$ with a sizable mass enhancement~\cite{Matsubayashi12,Matsubayashi14}.
The behavior of the characteristic energy resembles that shown in Fig.~\ref{flnfl}(b) if the pressure effect is interpreted as an appropriate change of parameters.
In this case, we have emphasized the importance of the orbital fluctuation between $\Gamma_{7}$ and $\Gamma_{8}$ in the vicinity of the FL-NFL boundary.
Such orbital fluctuation is also discussed in Ce-based compounds in the context of superconductivity with peculiar pressure dependence of $T_{c}$~\cite{Hattori10,Ren14}.
Therefore, it is highly desired to confirm the presence of orbital fluctuations by such as NQR measurement.
Meanwhile, the Kondo couplings for PrV$_{2}$Al$_{20}$ may be located in the NFL state of our model.
 
PrIr$_{2}$Zn$_{20}$ also shows clear NFL behavior in the resistivity~\cite{Onimaru10,Onimaru11,Ikeura14}.
The reduction of the entropy by applying magnetic field shows a close resemblance to that shown in Fig.~\ref{sh}(a)~\cite{Onimaru14}.
In high-field region above 6 T, the $T$ dependence of the resistivity shows a crossover from the NFL behavior to the ordinary FL one, and the specific-heat measurement confirms the splitting of the non-Kramers doublet.
There exists another crossover or transition at $T^{*}\sim 0.1$ K and $H\sim 5$ T, and its electronic state below $T^{*}$ remains unclear.
The most remarkable feature of the $H$ dependence of the resistivity is that it satisfies a scaling law~\cite{Machida14}.
Namely, the resistivity subtracted by the residual value as a function of $T/T_{\rm F2}^{*}$ for several values of $H$ falls into universal curve above $T_{\rm H}$ or $T_{\rm Q}$.
A scaling law for the specific heat under $H$ is also confirmed, in which $T_{\rm F2}^{*}$ is similar to that used in the case of the resistivity~\cite{Onimaru14}.
All of these observations provide strong evidence for the first realization of the (lattice) quadrupole Kondo effect and the resultant NFL behavior in 3-dimensional systems.
A metamagnetic behavior is observed around $H\sim 6$ T.
In our model, the metamagnetic behavior is associated with the change of the orbital occupation.
Thus, it would be interesting to confirm whether a change of the orbital occupation occurs or not such as by NQR measurements.

By applying the pressure to PrIr$_{2}$Zn$_{20}$, the resistivity for several pressures roughly follows the same scaling law as well~\cite{Umeo14}.
It also exhibits a crossover from the NFL to FL behavior above 8 GPa.
However, since its crossover temperature $T_{\rm FL}^{*}$ is always lower than $T_{\rm F2}^{*}$, there is a possibility of the dynamical Jahn-Tellar effect that lifts the degeneracy of the non-Kramers doublet.
Note that $T_{\rm F2}^{*}$ in this case also increases by pressure, which is similar to the case of PrTi$_{2}$Al$_{20}$~\cite{Umeo14,Matsubayashi14}.

In summary, we have discussed possible competition between the magnetic ($J$) and quadrupole ($K$) Kondo effects on the basis of the extended two-channel Kondo model.
There are three different states in the model, (i) the (local) Fermi-liquid state ($J\gg K$), (ii) the non-Fermi-liquid state ($J\ll K$), and (iii) the Fermi-liquid + marginally free-spin state ($J\approx K$), where (i) and (ii,iii) differ in the number of f electrons in the $\Gamma_{7}$ orbital.
Then, it is expected strong orbital fluctuation between $\Gamma_{7}$ and $\Gamma_{8}$ in the vicinity of the FL-NFL boundary.
The crossover temperature $T_{\rm H}$ and $T_{\rm 2K}$ under magnetic fields follows $\propto h^{2}$ in high-field region in contrast to the predicted $\propto h^{4}$ behavior in the low-field limit.
There is a possibility of the field-induced transition to the FL with metamagnetic behavior, provided that the parameters are close enough to the NFL-FL boundary.

These are general feature inherent from the $f^{2}$ non-Kramers doublet ground state with the moderate excited magnetic states under cubic symmetry.
Extensive investigation for corresponding dilute systems with respect to magnetic ions would be more informative in fundamental aspects of the competition between the magnetic and quadrupole Kondo effects.

\ack
The authors would like to thank K. Izawa, K. Umeo, K. Matsubayashi, and S. Nakatsuji for fruitful discussions and showing them the latest experimental data prior to publication.

\end{document}